\documentclass[PRX,preprint,amsmath,amssymb,superscriptaddress,longbibliography,endfloats*]{revtex4-1}

\usepackage{graphicx}
\usepackage{dcolumn}
\usepackage{bm}
\usepackage{hyperref}
\usepackage{breakurl}
\usepackage{multirow}
\usepackage{enumitem}
\usepackage{color}

\begin{document}

\title{Dialkali-Metal Monochalcogenide Semiconductors with High Mobility and Tunable Magnetism}

\author{Chenqiang Hua}
\affiliation{Department of Physics, Zhejiang University, Hangzhou 310027, P. R. China}
\affiliation{State Key Lab of Silicon Materials, School of Materials Science and Engineering, Zhejiang University, Hangzhou 310027, P. R. China}

\author{Feng Sheng}
\affiliation{Department of Physics, Zhejiang University, Hangzhou 310027, P. R. China}

\author{Qifeng Hu}
\affiliation{Department of Physics, Zhejiang University, Hangzhou 310027, P. R. China}

\author{Zhu-An Xu}
\affiliation{Department of Physics, Zhejiang University, Hangzhou 310027, P. R. China}
\affiliation{Collaborative Innovation Centre of Advanced Microstructures, Nanjing University, Nanjing 210093, P. R. China}

\author{Yunhao Lu}
\email{luyh@zju.edu.cn}
\affiliation
{State Key Lab of Silicon Materials, School of Materials Science and Engineering, Zhejiang University, Hangzhou 310027, P. R. China}

\author{Yi Zheng}
\email{phyzhengyi@zju.edu.cn}
\affiliation{Department of Physics, Zhejiang University, Hangzhou 310027, P. R. China}
\affiliation{Collaborative Innovation Centre of Advanced Microstructures, Nanjing University, Nanjing 210093, P. R. China}

\date{\today}

\begin{abstract}
The discovery of archetypal two-dimensional (2D) materials provides enormous opportunities in both fundamental breakthroughs and device applications, as evident by the research booming in graphene, atomically thin transition-metal chalcogenides, and few-layer black phosphorous in the past decade. Here, we report a new, large family of semiconducting dialkali-metal monochalcogenides (DMMCs) with an inherent A$_{2}$X monolayer structure, in which two alkali sub-monolayers form hexagonal close packing and sandwich the triangular chalcogen atomic plane. Such unique lattice structure leads to extraordinary physical properties, such as good dynamical and thermal stability, visible to near-infrared light energy gap, high electron mobility (e.g. $1.87\times10^{4}$ cm$^{2}$V$^{-1}$S$^{-1}$ in K$_{2}$O). Most strikingly, DMMC monolayers (MLs) host extended van Hove singularities near the valence band (VB) edge, which can be readily accessed by moderate hole doping of $\sim1.0\times10^{13}$ cm$^{-2}$. Once the critical points are reached, DMMC MLs undergo spontaneous ferromagnetic transition when the top VBs become fully spin-polarized by strong exchange interactions. Such gate tunable magnetism in DMMC MLs are promising for exploring novel device concepts in spintronics, electronics and optoelectronics.
\end{abstract} 

\maketitle

\section{Introduction}
Although the critical importance of dimensionality in determining the extraordinary physical properties of low-dimension systems has long been recognized since Richard Feyman, the groundbreaking experiments on graphene \cite{Gr04-Science-Geim,Gr05Geim_QHE,Gr05-QHE-KIM} provide a fascinating platform for exploring exotic phenomena in a rather simple hexagonal lattice. After the gold rush of graphene, the search of new two-dimensional (2D) systems have received unparalleled attention in the hope of novel physical properties and prototype functionalities. However, despite the flourishing of 2D materials \cite{NM-stanene15-JiaJF,PRL-Silicence12-LeLay,PRLSilicene-Ge-YaoYG,Va-Dirac-LuYH,Nature-Ca2N-Hosono,AFM-MXene13-Kawazoe,2Dreview15_ACSnano_Robinson}, the majority of 2D researches remain heavily focusing on the archetypal systems of graphene \cite{Geim13Nature_vdWCrystals}, transition-metal dichalcogenides (TMDCs) \cite{MoS2-PRL-Heinz,MoS2-FETmobility-NatNano-Kis,MoS2-Rippling-PRL15-ZY,Excitons-TMDCs-ROM18-Heinz}, and few-layer black phosphorus (BP) \cite{BP-FET-ZhangYB-NatNano,ZhangYB_16BPQHE_NatNano}. These 2D paradigms share the common features of weak interlayer coupling by van der Waals interactions, good thermal stability at room temperature, and high charge carrier mobility. Nevertheless, very recently, there are emergent 2D materials with tempting physical properties beyond the aforementioned systems, such as the InSe family \cite{GaSe_GateFM_PRL,InSe_QHE_NatNano}, 2D ferromagnetic van der Waals crystals \cite{Cr2Ge2Te6_2DFM_Nature,CrI3_2DFM_Nature}, binary maingroup compounds with the BP-type puckering lattice \cite{SnSe_BinaryBP_PRB,SnSe_Multiferro_NL,SnSe_SdH_NatCommun}, and the huge family of carbide and nitride based transition metal MXenes, in which M represents transition metal cation and X is C or N anion, respectively \cite{Nature-Ca2N-Hosono,ACSNano-Ti2C-Barsoum}.

Here, by using \textit{ab initio} density functional calculations, we discover a large family of dialkali-metal monochalcogenides (DMMCs) with excellent dynamical and thermal stability, visible to near-infrared light energy gap, and very high electron mobility of exceeding $1.0\times10^{3}\,\textrm{cm}^{2}\textrm{V}^{-1}\textrm{s}^{-1}$. With the same lattice structure of 1T-TMDCs, DMMCs have the inherent layer-by-layer structure with very weak interlayer coupling, due to the intrinsic +1 oxidation state of alkali cations. All these features make DMMCs a unique choice beyond TMDCs for fundamental studies of 2D systems and for developing potential electronic and optoelectronic devices. Most fascinatingly, all DMMC monolayers (MLs) host extended singularity points in the density of states near the Mexican-hat shaped valence band maximum (VBM). Using hole doping in the range of $\sim 1.0\times10^{13}- 2.5\times10^{13}$ cm$^{-2}$ via electrostatic or liquid ion gating, spontaneous ferromagnetic transitions, which lead to half-metallic character with full spin polarized top valence bands, can be triggered by strong exchange interactions in these 2D systems. Such gate tunable magnetism and half metallicity in DMMC MLs may pave new routes in novel device concepts for spintronics.

\section{Results and Discussion}
\subsection{Structure and Stability}
\begin{figure*}[!thb]
\includegraphics[scale=1]{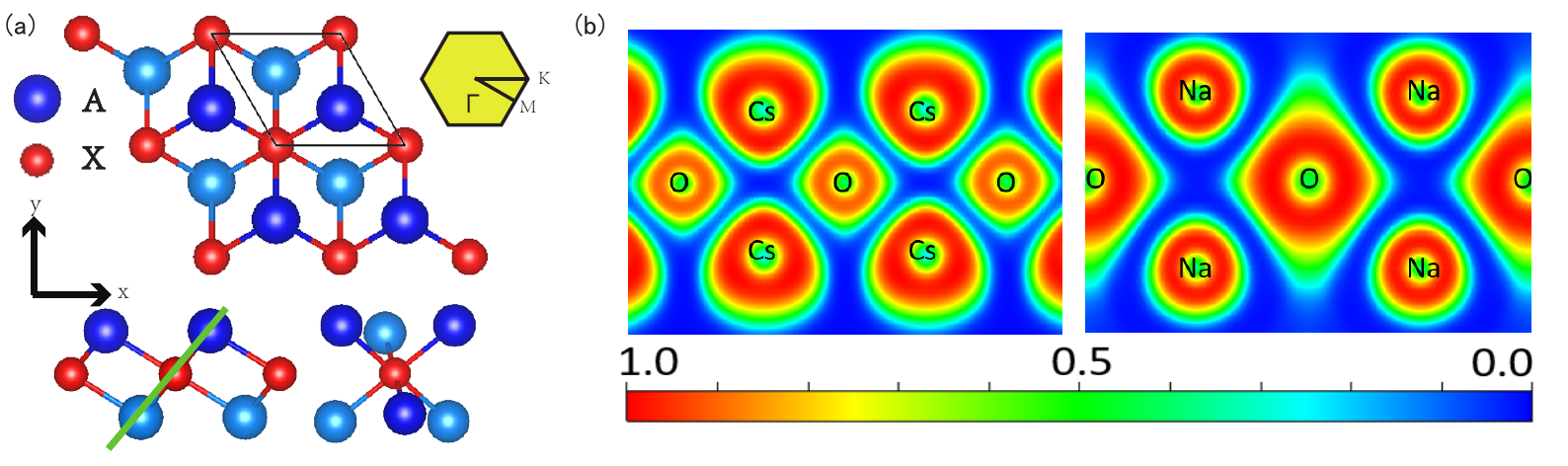}
\caption{(\textbf{a}) Structure of DMMC MLs. The upper inset is the first Brillouin zone.  Each chalcogen anion is surrounded by six alkali cations, forming XA$_{6}$ octahedron. (\textbf{b-c}) Normalized electron localization function (ELF) of Cs$_{2}$O and Na$_{2}$O, respectively. ELF = 1 (red) and 0 (blue) indicate accumulated and vanishing electron densities, respectively. The 2D ELF plane is defined by the green line in Fig. 1a. }
\label{fig1}
\end{figure*}

\begin{figure*}[!thb]
\includegraphics[scale=1]{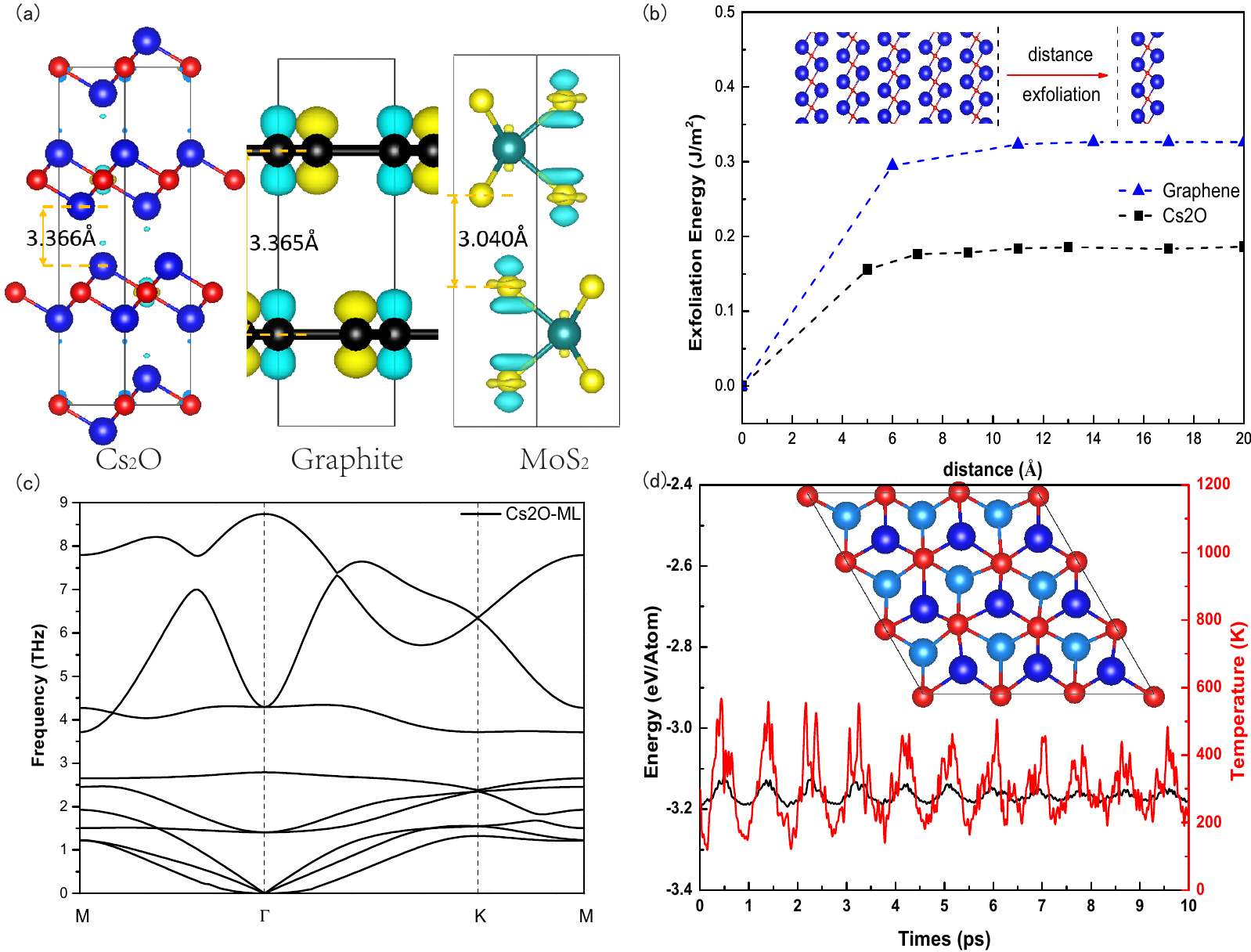}
\caption{\textbf{Stability and Micromechanical Exfoliation.} (\textbf{a}) Interlayer differential charge density of Cs$_{2}$O bulk, graphite and MoS$_{2}$ bulk. (\textbf{b}) Calculated exfoliation energy of Cs$_{2}$O ML, in comparison with graphene (blue triangles). (\textbf{c}) The phonon spectrum of Cs$_{2}$O ML. (\textbf{d}) First-principle molecular dynamics simulations of Cs$_{2}$O ML. The thermal stability of the supercell is evident by the convergence in the free energy. }
\label{fig2}
\end{figure*}

Dialkali-metal monochalcogenides form a large family with sixteen compounds in the general formula of A$_{2}$X, where A represents an alkali atom (Na, K, Rb or Cs) and X is a chalcogen anion of O, S, Se or Te. The synthesis of bulk dicesium monoxide (Cs$_{2}$O) was first reported in 1955 \cite{Cs2Oexp_Lassettre_JPC} with the space group $R\bar{3}m$ (No. 166), which has the same lattice structures as the well-known 1T-TMDCs \cite{TMDCsReview_AdvPhys} as displayed in Fig. \ref{fig1}a and \ref{fig2}a. By taking van der Waals (vdW) corrections into account, the optimized lattice constants of bulk Cs$_{2}$O are $a=4.23$ \AA, $b=4.23$ \AA, and $c=19.88$ \AA, respectively, which are in excellent agreement with the experimental values \cite{Cs2Oexp_Lassettre_JPC}. In the 2D limit, the monolayer of DMMCs has the space group $P\bar{3}m1$ (No. 164). The primitive unit cell consists of two alkali cations and one group-VIa anions, making A$_{2}$X stoichiometry. The metal-chalcogen ratio in DMMC MLs is distinct from the well-known TMDC MLs, in which the unit cell includes two chalcogenide anions and one metal cation. For instance, in Cs$_{2}$O ML, the centering O atom is surrounded by six Cs cations, forming a distinctive O-Cs octahedron, as shown in Fig. \ref{fig1}a. Such lattice structure is rooted in strong intralayer O-Cs ionic bonding, as revealed by the analysis of electron localization functions (ELFs). As shown in Fig. \ref{fig1}b, the electron densities are highly localized around Cs and O atoms with negligible inter-atom distribution, reflecting ionic bonding and electron donation from Cs to O atoms. The calculated oxidation state of O anions, as represented by the Hirshfeld charge, is nearly identical to that of BaO, also confirming the dianionic character of oxygen in Cs$_{2}$O. No accumulation of electron density is observed between Cs atoms, suggesting no chemical bond between them. The A-X octahedral geometry centring the chalcogen dianion is general observed for the whole family, as presented in Fig. \ref{fig1}b for Na$_{2}$O ML and in Supplementary Information (SI) Figure S1-S4 for the other 14 DMMC MLs.

Due to the intrinsic full oxidation state (+1) of Cs, the interlayer coupling between Cs sub-MLs is very weak, which only introduces marginal changes to the in-plane lattice constants. The optimized $a$ and $b$ of Cs$_{2}$O ML (4.26 \AA) are almost the same as the experimental bulk value. Comparing with the paradigmatic 2D systems of MoS$_{2}$ and graphite, we find that there are even less interlayer electron density localization in Cs$_{2}$O as a result of dominant in-plane ionic bonding, which also means micromechanical exfoliation of Cs$_{2}$O crystals can be readily achieved (Fig. \ref{fig2}a). Indeed, we calculate the exfoliation energy of Cs$_{2}$O ML to be  $\sim 0.19$ J/m$^{2}$ (Fig. \ref{fig2}b), which is surprisingly lower than the value of 0.33 J/m$^{2}$ for graphene \cite{Gr-Cleavage-PRB}. It has to been emphasized here that the lower exfoliation energy of bulk Cs$_{2}$O, compared with MoS$_{2}$ ($\sim$ 0.42 J/m$^{2}$) and graphite, is consistently obtained by different vdW correction methods, which is summarized in SI Table S1.

To assess the dynamical stability, which is also crucial for the micro-exfoliation of Cs$_{2}$O ML, we have computed the phonon dispersion with the finite displacement method. As shown in Fig. \ref{fig2}c, the good stability of ML Cs$_{2}$O is evident by the positive values of all phonon modes. We further check the room-temperature (RT) thermal stability of Cs$_{2}$O ML by first-principle molecular dynamics simulations. The fluctuation of energy and temperature as a function of time are plotted in Fig. \ref{fig2}d. After running 4000 steps (10 ps), the 3$\times$3$\times$1 trigonal lattice is well sustained, and the free energy of the supercell converges. Excellent dynamical and thermal stability have also been validated for the other DMMC MLs (See SI Figure S5 and S6). Thus, it is feasible to experimentally exfoliate DMMC MLs for device fabrications and applications.

\subsection{Electronic Properties}
\begin{figure*}[!thb]
\includegraphics[scale=1]{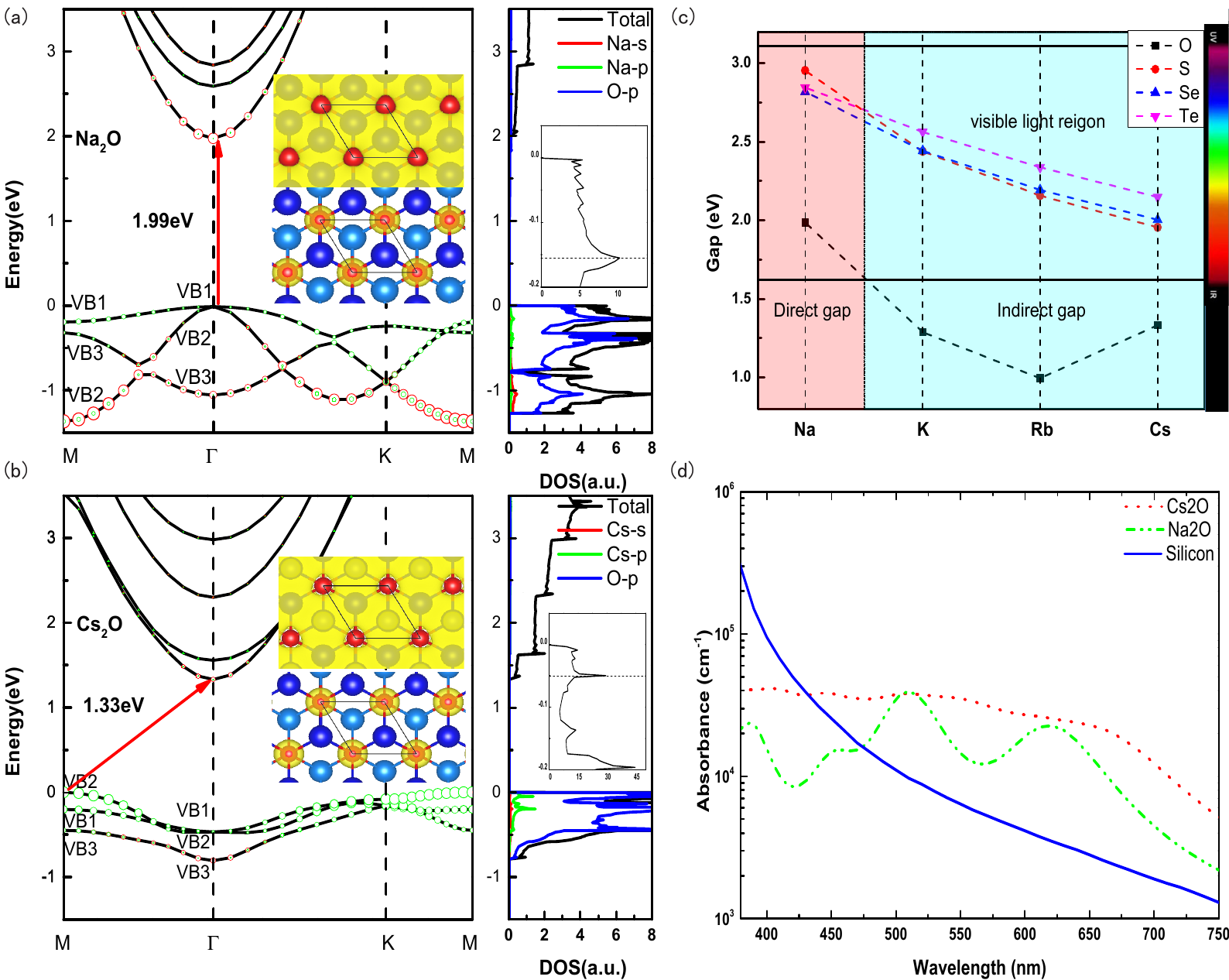}
\caption{\textbf{Electronic Properties.} Band structures and the corresponding DOS of Na$_{2}$O ML (\textbf{a}) and Cs$_{2}$O ML (\textbf{b}). The contributions of alkali $s$-orbital and $p$-orbital are indicated by red and green circles, respectively. The insets are the spatial distribution of the wave-functions for the VBM and CBM. (\textbf{c}) Energy gaps of all DMMC MLs. Strong inter $p$-orbital hybridization leads to the transition from direct bandgaps in Na$_{2}$X to indirect gaps in other DMMC MLs. The gap values of DMMC MLs are mainly determined by the electron affinity of the anions. The results have been confirmed by both the HSE06 and PBE methods. (\textbf{d}) Extraordinary light absorbance of Cs$_{2}$O and Na$_{2}$O MLs, in comparison with silicon.}
\label{fig3}
\end{figure*}

We now elucidate the electronic structures and fundamental physical properties of DMMC MLs using two representative examples of Cs$_{2}$O and Na$_{2}$O MLs. As shown in Fig. \ref{fig3}a, Na$_{2}$O ML is a 2D semiconductor with a direct bandgap of $\sim$ 1.99 eV at the $\Gamma$ point. Noticeably, the VBM of Na$_{2}$O ML is mainly contributed by O-2$p$ orbitals (its projection weight is not shown in the energy dispersion for clarity) hybridized with very low weight of Na-2$p$ (weighted by green triangles), while the CBM is dominated by Na-3$s$ (red squares). The electronic band structure undergoes a pronounced change when Na is replaced by Cs, whose strongly delocalized 5$p$ orbitals induce significant density of states (DOS) away from the $\Gamma$ point, which is still dominated by O-2$p$. As the energy of Cs-5$p$ orbitals is $\sim 0.4$ eV higher than the 2$p$ orbitals of O, the inter $p$-orbital hybridization reverses the energy levels of VB1 and VB2 along the $\Gamma-M$ direction. The resulting Cs$_{2}$O ML is an indirect semiconductor with multiple VBMs centering the $M$ points (Fig. \ref{fig3}b). In general, for heavier cations of K, Rb and Cs, the inter $p$-orbital hybridization with chalcogen anions is prevailing, shifting VBMs from $\Gamma$ to the $M$ point. As a result, except for Na$_{2}$X (X=O, S, Se and Te), all DMMC MLs are indirect semiconductors. The inter $p$-orbital hybridization induced direct to indirect bandgap transition from Na$_{2}$X to other A$_{2}$X MLs has also been confirmed by the PBE method, as shown in SI Table S2 and SI Figure S7.

As summarized in Fig. \ref{fig3}c, the energy gaps of DMMC MLs range from 1 eV to 3 eV, covering the near-infrared (K$_{2}$O, Rb$_{2}$O and Cs$_{2}$O MLs) and visible light regions, which are highly desirable for optoelectronic device applications. To evaluate the optical performance, the absorption coefficients of Cs$_{2}$O and Na$_{2}$O MLs are directly compared with bulk silicon \cite{Si-absorb-exp}. As shown in Fig. \ref{fig3}d, Na$_{2}$O ML shows absorption attenuation above $\sim 625$ nm, in consistent with the direct bandgap of 1.99 eV. Strikingly, indirect-bandgap Cs$_{2}$O ML exhibits unrivaled cyan-to-red light absorption efficiency, which is nearly constant over the whole visible light wavelengths. Such unusual findings not only predict Cs$_{2}$O MLs to be extraordinary optoelectronic materials, but also are rooted in the enormous, non-differentiable DOS in the vicinity of VBM, \textit{i.e.} van Hove singularity. As shown in Fig. \ref{fig3}a and \ref{fig3}b, the top VBs of DMMC MLs, dominated by the localized $p$ orbitals of chalcogen anions, are distinctive by very flat dispersion in the momentum space. The inter $p$-orbital hybridization in Cs$_{2}$O ML drastically reduces the VB energy dispersion, which effectively enhances DOS near the VBM and creates multiple saddle points in the vicinity of $\Gamma$, $K$ and $M$. This leads to a prominent extended van Hove singularity in the density of states \cite{YBCO-vanHove_PRL94}. It is also noteworthy that the inter $p$-orbital hybridization also greatly enhances DOS near the CBM, which contributes significantly to the extraordinary light absorption characteristics of Cs$_{2}$O ML. 

Charge carrier mobility is another critical parameter for the device performance of 2D material-based devices. As summarized in Table \ref{tb1} as well as in SI Table S3 and S4, the mobility of DMMC MLs are largely asymmetric between electrons and holes due to the drastic difference in the energy dispersion between the CBs and VBs, producing highly asymmetric effective mass. Noticeably, K$_{2}$O ML has the highest electron mobility of 1.87$\times10^{4}$ cm$^{2}$V$^{-1}$S$^{-1}$ along the $y$ direction. Such electron mobility is comparable to the hole mobility of few-layer BP \cite{BP-calculation-NC-JiW,ZhangYB_16BPQHE_NatNano} and order of magnitude larger than that of MoS${_2}$ atomic layers \cite{MoS2-FETmobility-NatNano-Kis}. Large electron mobility in DMMC MLs may hold a great promise for applications in high-performance electronics.

We have noticed that with the same trigonal structure ($R\bar{3}m$, No.166), monolayer Tl$_{2}$O is predicted to be a semiconductor with a direct bandgap and highly anisotropic charge carrier mobilities up to $4.3\times10^{3}\,\textrm{cm}^{2}\textrm{V}^{-1}\textrm{s}^{-1}$ \cite{Tl2O_JACS17_Heine}. This is not surprising since Tl is one of the pseudo-alkali metals, in close resemblance to K both in the oxidation state and ionic radius.

\subsection{Tunable Magnetism by Hole doping.}
\begin{figure*}[!thb]
\includegraphics[scale=1]{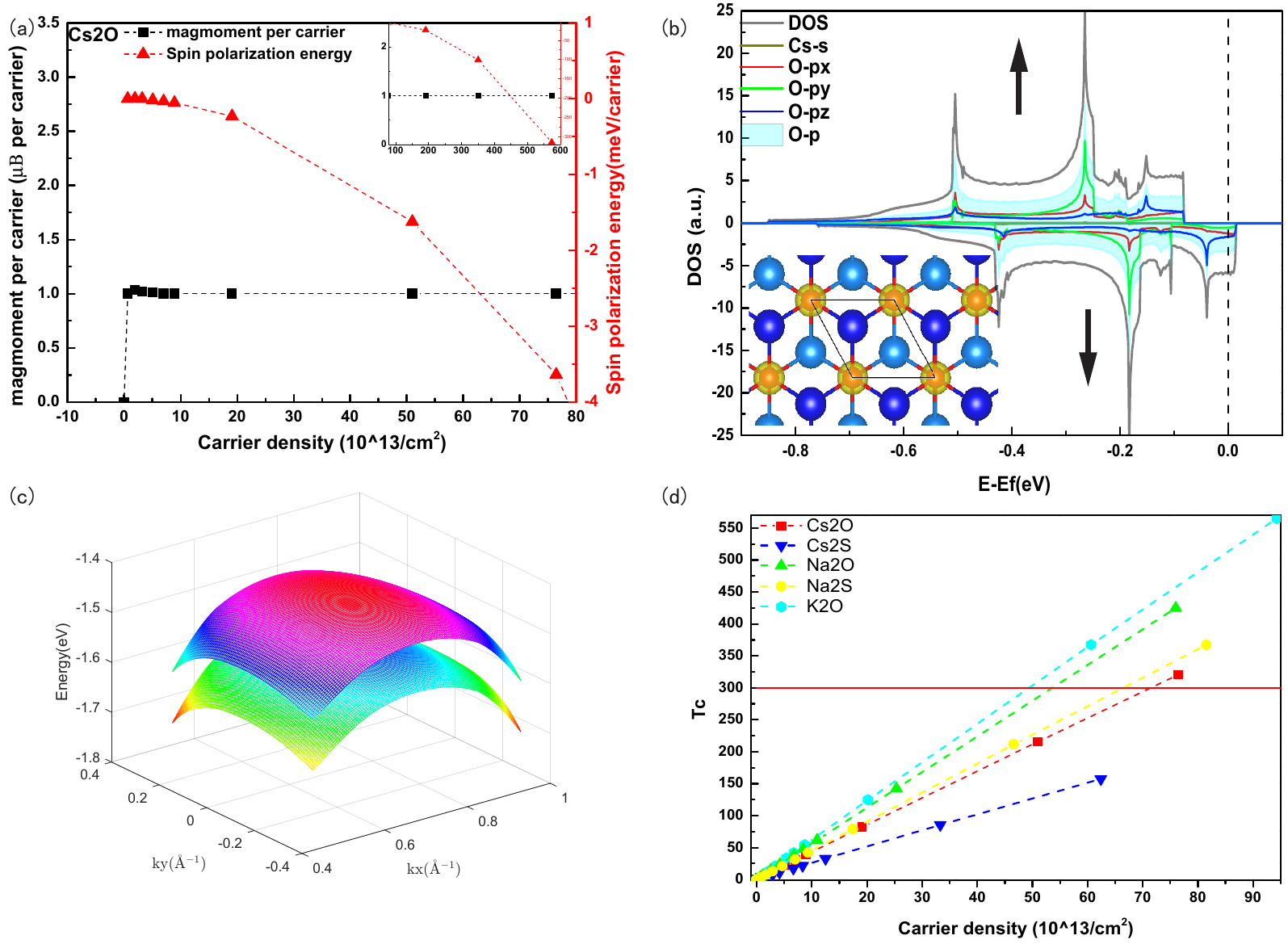}
\caption{\textbf{Tunable Magnetism by Hole Doping.} (\textbf{a}) Magnetic moment and spin polarization energy per carrier of Cs$_{2}$O ML. Remarkably, spontaneous ferromagnetic phase transition is trigger by a moderate hole doping of $1.28\times10^{13}$ cm$^{-2}$. (\textbf{b}) Spin-resolved DOS of Cs$_{2}$O ML at $n=5\times10^{14}$ cm$^{-2}$. The inset shows the spin density distribution, apparently locating at O atoms. (\textbf{c}) Spin-splitted top valence bands in the vicinity of $M$ points in Cs$_{2}$O ML ($n=5\times10^{14}$ cm$^{-2}$). (\textbf{d}) Curie temperature $verus$ hole doping density, showing the feasibility of RT ferromagnetism by heavy hole doping. Note that antiferromagnetic ordering in DMMC MLs is energetically higher in free energy than the ferromagnetic state.}
\label{fig4}
\end{figure*}

The aforementioned van Hove singularities in the top VB of DMMC MLs are associated with inherent electronic instability, when hole doping pushes the Fermi level approaching the divergent point. Taking Cs$_{2}$O as an example, we find that the system undergoes a spontaneous ferromagnetic phase transition with a moderate hole doping of $n\sim1.0\times10^{13}$ cm$^{-2}$. In Fig. \ref{fig4}a, we plot the magnetic moment and the spin polarization energy per doped hole carrier as a function of $n$. The latter represents the energy difference between the non-spin-polarized phase and the ferromagnetic phase. Clearly, the top VB becomes fully spin polarized by the critical hole doping, as manifested by the constant magnetic moment of 1 $\mu_{B}$ per carrier above $1.0\times10^{13}$ cm$^{-2}$. The spin polarization energy, which becomes -3.11 meV per carrier for a minimal hoel doping, monotonically decreases as a function of $n$, approving that the doping-induced ferromagnetic ordering is energy favourable. 

Such spontaneous ferromagnetic ordering can be well understood by the Stoner model, which has been adopted to explain the so-called ``$d^{0}$ ferromagnetism'' in hole doped nitrides and oxides \cite{ZnO-FM-PRL}. In this picture, ferromagnetism can spontaneously appear when the Stoner criterion $U*g(E_{f})>$1 is satisfied, where $U$ is the exchange interaction strength and $g(E_{f})$ is the DOS at the Fermi energy of non-magnetic state. In such condition, the energy gain by exchange interactions exceeds the loss in kinetic energy, and hence the system would favor a ferromagnetic ground state. We have confirmed that the $U*g(E_{f})$ of Cs$_{2}$O is always larger than one above the critical doping level, indicating that the huge DOS associated with the van Hove singularity plays a key role in the occurrence of ferromagnetism.

In Fig. \ref{fig4}b and \ref{fig4}c, we plot the spin-polarized DOS of Cs$_{2}$O with a doping level of $n=5.1\times10^{14}$ cm$^{-2}$. Compared with the undoped case (Fig. \ref{fig3}b), the corresponding spin splitting process causes a significant energy shit of 0.1 eV between the spin-up and spin-down bands, leaving the Fermi level cutting through only one spin channel. Such half-metal state allows fully spin-polarized transport, which is crucial for spintronics applications. It is fascinating that this half-metallic ferromagnetism can be reversibly switched on and off in a $\sim$1 nm thick atomic layer without extrinsic dopant and defects. 

We have found that except for Na$_{2}$Te, the ferromagnetic transition can be triggered in all DMMC MLs by hole doping, ranging from $\sim1.0\times10^{13}$ cm$^{-2}$ for Cs$_{2}$O to $8.0\times10^{15}$ cm$^{-2}$ for Na$_{2}$Se (see SI Figure S8-S11). The special case of Na$_{2}$Te can be understood by insufficient DOS near the VBM, which is dominated by the 5$p$ orbitals of Te. Due to much lower electron negativity, the 5$p$ orbitals of Te are more delocalized than the other chalcogen elements. Consequently, the high dispersion of Te-5$p$ reduce DOS near the VBM of Na$_{2}$Te (SI Fig. S8), invalidating the Stoner criterion. We further evaluated the doping dependent Curie temperature ($T_{c}$) of DMMC MLs using the Heisenberg model. As shown in Fig. \ref{fig4}d, using $n=5.0\times10^{14}$ cm$^{-2}$ as a reference, which can be readily induced by ionic liquid gating or lithium glass gating \cite{BP-FET-ZhangYB-NatNano,FeSegate2016-PRL-ChenXH}, RT ferromagnetism is achievable in K$_{2}$O and Na$_{2}$O MLs (SI Figure S12). For Cs$_{2}$O, the corresponding $T_{c}$ is $\sim 200$ K, which is nearly five times higher than the recently reported ferromagnetic CrI$_{3}$ and Cr$_{2}$Ge$_{2}$Te$_{6}$ \cite{Cr2Ge2Te6_2DFM_Nature,CrI3_2DFM_Nature}.

\section{Conclusions}
In summary, we have reported a new family of 1T-TMDC structured dialkali-metal monochalcogenides (DMMCs), which show fascinating physical properties and great device application promises in the 2D limit. Distinctively, DMMC MLs are characterized by large electron mobility, e.g. $1.87\times10^{4}$ cm$^{2}$V$^{-1}$S$^{-1}$ for K$_{2}$O and $5.42\times10^{3}$ cm$^{2}$V$^{-1}$S$^{-1}$ for Na$_{2}$O, and by very flat VBs with rather weak energy dispersion. The latter are responsible for the formation of extended van Hove singularities with extremely high DOS just below the VB edge of DMMC MLs. By introducing moderate hole doping, these DMMC MLs are subjected to strong electronic instabilities induced by electron-electron interaction, which ultimately leads to spontaneous ferromagnetic phase transitions where the top VBs becomes fully spin-polarized. Using heavy doping methods of ion liquid and solid gating, we can readily increase the $T_{c}$ and switch on and off the magnetism of DMMC MLs even at room temperature. 

Although alkali compounds are not very stable in air by reacting with ambient moisture, the excellent dynamical and thermal stability allows DMMC MLs to be exfoliated and encapsulated in glove box environment, which is now routine research facilities for studying air-sensitive 2D materials \cite{BP-FET-ZhangYB-NatNano,ZhangYB_16BPQHE_NatNano,InSe_QHE_NatNano,Cr2Ge2Te6_2DFM_Nature,CrI3_2DFM_Nature}. It is also feasible to grow DMMC MLs by molecular beam epitaxy or by chemical vapor deposition, with abundant options of single crystal substrates of 1T- or 2H-TMDCs. By forming heterostructure with high electron affinity main-group MDCs, such as SnSe$_{2}$, DMMC MLs may become effectively hole doped by interfacial charge transferring \cite{SnSe_SdH_NatCommun}, which would provide the opportunity for in-situ studying the gate-tunable ferromagnetism in DMMC MLs by scanning probe microscope.


\begin{table*}
 \caption{\textbf{Mobility and effective mass along the $x$ and $y$ transport directions in four Cs$_{2}$X MLs and K$_{2}$O}. We calculated the carrier mobility using a phonon-limited scattering model, which includes deformation potential (E$_{1}$) and elastic modulus (C$_{2D}$) in the propagation direction of the longitudinal acoustic wave. Noticeably, K$_{2}$O has the highest electron mobility.}
 \label{tb1}
 \begin{tabular}{ccccc}
 \hline

Effective mass (m$_{0}$) &  m$_{hx}$&   m$_{hy}$ &   m$_{ex}$  &  m$ _{ey}$  \\
\hline

Cs$_{2}$O &5.36  &4.71  &0.63 &0.84                 \\
Cs$_{2}$S &1.02  &2.13  &0.67 &0.89                 \\
Cs$_{2}$Se &0.95  &1.99  &0.61 &0.81                 \\
Cs$_{2}$Te &1.03  &1.88  &0.58 &0.77                \\
K$_{2}$O  &1.84  &11.2  &0.90  &0.67             \\
\hline
Mobility (cm$^{2}$V$^{-1}$S$^{-1}$) & $\mu_{hx}$   &  $\mu_{hy}$   &   $\mu_{ex}$& $\mu_{ey}$ \\
\hline
Cs$_{2}$O&    $0.13\times10^{1}$ & $0.34\times10^{1}$ & $9.63\times10^{1}$ & $1.87\times10^{1}$\\
Cs$_{2}$S&    $0.29\times10^{1}$ & $0.36\times10^{1}$ & $6.99\times10^{1}$ & $1.10\times10^{2}$\\
Cs$_{2}$Se&    $0.15\times10^{1}$ & $9.51\times10^{-1}$ & $5.60\times10^{1}$ & $4.33\times10^{1}$\\
Cs$_{2}$Te&    $9.07\times10^{-1}$ & $6.95\times10^{-1}$ & $2.61\times10^{1}$ & $2.37\times10^{1}$\\
K$_{2}$O&$0.46\times10^{1}$  & $2.53\times10^{-1}$  & $\mathbf{8.80\times10^{3}}$ & $\mathbf{1.87\times10^{4}}$\\

\hline
 \end{tabular}
\end{table*}

\begin{acknowledgments} This work is supported by the National Key R\&D Program of the MOST of China (Grant Nos. 2016YFA0300204 and 2017YFA0303002), and the National Science Foundation of China (Grant Nos. 11574264 and 61574123). Y.Z. acknowledges the funding support from the Fundamental Research Funds for the Central Universities and the Thousand Talents Plan.
\end{acknowledgments}


\section*{Competing financial interests:}
The authors declare no competing financial interests.


\begin{thebibliography}{34}%
\makeatletter
\providecommand \@ifxundefined [1]{%
 \@ifx{#1\undefined}
}%
\providecommand \@ifnum [1]{%
 \ifnum #1\expandafter \@firstoftwo
 \else \expandafter \@secondoftwo
 \fi
}%
\providecommand \@ifx [1]{%
 \ifx #1\expandafter \@firstoftwo
 \else \expandafter \@secondoftwo
 \fi
}%
\providecommand \natexlab [1]{#1}%
\providecommand \enquote  [1]{``#1''}%
\providecommand \bibnamefont  [1]{#1}%
\providecommand \bibfnamefont [1]{#1}%
\providecommand \citenamefont [1]{#1}%
\providecommand \href@noop [0]{\@secondoftwo}%
\providecommand \href [0]{\begingroup \@sanitize@url \@href}%
\providecommand \@href[1]{\@@startlink{#1}\@@href}%
\providecommand \@@href[1]{\endgroup#1\@@endlink}%
\providecommand \@sanitize@url [0]{\catcode `\\12\catcode `\$12\catcode
  `\&12\catcode `\#12\catcode `\^12\catcode `\_12\catcode `\%12\relax}%
\providecommand \@@startlink[1]{}%
\providecommand \@@endlink[0]{}%
\providecommand \url  [0]{\begingroup\@sanitize@url \@url }%
\providecommand \@url [1]{\endgroup\@href {#1}{\urlprefix }}%
\providecommand \urlprefix  [0]{URL }%
\providecommand \Eprint [0]{\href }%
\providecommand \doibase [0]{http://dx.doi.org/}%
\providecommand \selectlanguage [0]{\@gobble}%
\providecommand \bibinfo  [0]{\@secondoftwo}%
\providecommand \bibfield  [0]{\@secondoftwo}%
\providecommand \translation [1]{[#1]}%
\providecommand \BibitemOpen [0]{}%
\providecommand \bibitemStop [0]{}%
\providecommand \bibitemNoStop [0]{.\EOS\space}%
\providecommand \EOS [0]{\spacefactor3000\relax}%
\providecommand \BibitemShut  [1]{\csname bibitem#1\endcsname}%
\let\auto@bib@innerbib\@empty
\bibitem [{\citenamefont {Novoselov}\ \emph {et~al.}(2004)\citenamefont
  {Novoselov}, \citenamefont {Geim}, \citenamefont {Morozov}, \citenamefont
  {Jiang}, \citenamefont {Zhang}, \citenamefont {Dubonos}, \citenamefont
  {Grigorieva},\ and\ \citenamefont {Firsov}}]{Gr04-Science-Geim}%
  \BibitemOpen
  \bibfield  {author} {\bibinfo {author} {\bibfnamefont {K.~S.}\ \bibnamefont
  {Novoselov}}, \bibinfo {author} {\bibfnamefont {A.~K.}\ \bibnamefont {Geim}},
  \bibinfo {author} {\bibfnamefont {S.~V.}\ \bibnamefont {Morozov}}, \bibinfo
  {author} {\bibfnamefont {D.}~\bibnamefont {Jiang}}, \bibinfo {author}
  {\bibfnamefont {Y.}~\bibnamefont {Zhang}}, \bibinfo {author} {\bibfnamefont
  {S.~V.}\ \bibnamefont {Dubonos}}, \bibinfo {author} {\bibfnamefont {I.~V.}\
  \bibnamefont {Grigorieva}}, \ and\ \bibinfo {author} {\bibfnamefont {A.~A.}\
  \bibnamefont {Firsov}},\ }\bibfield  {title} {\enquote {\bibinfo {title}
  {Electric field effect in atomically thin carbon films},}\ }\href {\doibase
  10.1126/science.1102896} {\bibfield  {journal} {\bibinfo  {journal}
  {Science}\ }\textbf {\bibinfo {volume} {306}},\ \bibinfo {pages} {666--669}
  (\bibinfo {year} {2004})}\BibitemShut {NoStop}%
\bibitem [{\citenamefont {Novoselov}\ \emph {et~al.}(2005)\citenamefont
  {Novoselov}, \citenamefont {Geim}, \citenamefont {Morozov}, \citenamefont
  {Jiang}, \citenamefont {Katsnelson}, \citenamefont {Grigorieva},
  \citenamefont {Dubonos},\ and\ \citenamefont {Firsov}}]{Gr05Geim_QHE}%
  \BibitemOpen
  \bibfield  {author} {\bibinfo {author} {\bibfnamefont {K.~S.}\ \bibnamefont
  {Novoselov}}, \bibinfo {author} {\bibfnamefont {A.~K.}\ \bibnamefont {Geim}},
  \bibinfo {author} {\bibfnamefont {S.~V.}\ \bibnamefont {Morozov}}, \bibinfo
  {author} {\bibfnamefont {D.}~\bibnamefont {Jiang}}, \bibinfo {author}
  {\bibfnamefont {M.~I.}\ \bibnamefont {Katsnelson}}, \bibinfo {author}
  {\bibfnamefont {I.~V.}\ \bibnamefont {Grigorieva}}, \bibinfo {author}
  {\bibfnamefont {S.~V.}\ \bibnamefont {Dubonos}}, \ and\ \bibinfo {author}
  {\bibfnamefont {A.~A.}\ \bibnamefont {Firsov}},\ }\bibfield  {title}
  {\enquote {\bibinfo {title} {Two-dimensional gas of massless {D}irac fermions
  in graphene},}\ }\href {\doibase 10.1038/nature04233} {\bibfield  {journal}
  {\bibinfo  {journal} {Nature}\ }\textbf {\bibinfo {volume} {438}},\ \bibinfo
  {pages} {197--200} (\bibinfo {year} {2005})}\BibitemShut {NoStop}%
\bibitem [{\citenamefont {Zhang}\ \emph {et~al.}(2005)\citenamefont {Zhang},
  \citenamefont {Tan}, \citenamefont {Stormer},\ and\ \citenamefont
  {Kim}}]{Gr05-QHE-KIM}%
  \BibitemOpen
  \bibfield  {author} {\bibinfo {author} {\bibfnamefont {Y.~B.}\ \bibnamefont
  {Zhang}}, \bibinfo {author} {\bibfnamefont {Y.~W.}\ \bibnamefont {Tan}},
  \bibinfo {author} {\bibfnamefont {H.~L.}\ \bibnamefont {Stormer}}, \ and\
  \bibinfo {author} {\bibfnamefont {P.}~\bibnamefont {Kim}},\ }\bibfield
  {title} {\enquote {\bibinfo {title} {Experimental observation of the quantum
  {H}all effect and {B}erry's phase in graphene},}\ }\href {\doibase
  10.1038/nature04235} {\bibfield  {journal} {\bibinfo  {journal} {Nature}\
  }\textbf {\bibinfo {volume} {438}},\ \bibinfo {pages} {201--204} (\bibinfo
  {year} {2005})}\BibitemShut {NoStop}%
\bibitem [{\citenamefont {Zhu}\ \emph {et~al.}(2015)\citenamefont {Zhu},
  \citenamefont {Chen}, \citenamefont {Xu}, \citenamefont {Gao}, \citenamefont
  {Guan}, , \citenamefont {Liu}, \citenamefont {Qian}, \citenamefont {Zhang},\
  and\ \citenamefont {Jia}}]{NM-stanene15-JiaJF}%
  \BibitemOpen
  \bibfield  {author} {\bibinfo {author} {\bibfnamefont {F.}~\bibnamefont
  {Zhu}}, \bibinfo {author} {\bibfnamefont {W.}~\bibnamefont {Chen}}, \bibinfo
  {author} {\bibfnamefont {Y.}~\bibnamefont {Xu}}, \bibinfo {author}
  {\bibfnamefont {C.}~\bibnamefont {Gao}}, \bibinfo {author} {\bibfnamefont
  {D.}~\bibnamefont {Guan}}, , \bibinfo {author} {\bibfnamefont
  {C.}~\bibnamefont {Liu}}, \bibinfo {author} {\bibfnamefont {D.}~\bibnamefont
  {Qian}}, \bibinfo {author} {\bibfnamefont {S.~C.}\ \bibnamefont {Zhang}}, \
  and\ \bibinfo {author} {\bibfnamefont {J.~F.}\ \bibnamefont {Jia}},\
  }\bibfield  {title} {\enquote {\bibinfo {title} {Epitaxial growth of
  two-dimensional stanene},}\ }\href {\doibase 10.1038/nmat4384} {\bibfield
  {journal} {\bibinfo  {journal} {Nature Mater.}\ }\textbf {\bibinfo {volume}
  {14}},\ \bibinfo {pages} {1020--1025} (\bibinfo {year} {2015})}\BibitemShut
  {NoStop}%
\bibitem [{\citenamefont {Vogt}\ \emph {et~al.}(2012)\citenamefont {Vogt},
  \citenamefont {De~Padova}, \citenamefont {Quaresima}, \citenamefont {Avila},
  \citenamefont {Frantzeskakis}, , \citenamefont {Asensio}, \citenamefont
  {Resta}, \citenamefont {Ealet},\ and\ \citenamefont
  {Le~Lay}}]{PRL-Silicence12-LeLay}%
  \BibitemOpen
  \bibfield  {author} {\bibinfo {author} {\bibfnamefont {P.}~\bibnamefont
  {Vogt}}, \bibinfo {author} {\bibfnamefont {P.}~\bibnamefont {De~Padova}},
  \bibinfo {author} {\bibfnamefont {C.}~\bibnamefont {Quaresima}}, \bibinfo
  {author} {\bibfnamefont {J.}~\bibnamefont {Avila}}, \bibinfo {author}
  {\bibfnamefont {E.}~\bibnamefont {Frantzeskakis}}, , \bibinfo {author}
  {\bibfnamefont {M.~C.}\ \bibnamefont {Asensio}}, \bibinfo {author}
  {\bibfnamefont {A.}~\bibnamefont {Resta}}, \bibinfo {author} {\bibfnamefont
  {B.}~\bibnamefont {Ealet}}, \ and\ \bibinfo {author} {\bibfnamefont
  {G.}~\bibnamefont {Le~Lay}},\ }\bibfield  {title} {\enquote {\bibinfo {title}
  {Silicene: Compelling experimental evidence for graphenelike two-dimensional
  silicon},}\ }\href {\doibase 10.1103/PhysRevLett.108.155501} {\bibfield
  {journal} {\bibinfo  {journal} {Phys. Rev. Lett.}\ }\textbf {\bibinfo
  {volume} {108}},\ \bibinfo {pages} {155501} (\bibinfo {year}
  {2012})}\BibitemShut {NoStop}%
\bibitem [{\citenamefont {Liu}\ \emph {et~al.}(2011)\citenamefont {Liu},
  \citenamefont {Feng},\ and\ \citenamefont {Yao}}]{PRLSilicene-Ge-YaoYG}%
  \BibitemOpen
  \bibfield  {author} {\bibinfo {author} {\bibfnamefont {C.~C.}\ \bibnamefont
  {Liu}}, \bibinfo {author} {\bibfnamefont {W.~X.}\ \bibnamefont {Feng}}, \
  and\ \bibinfo {author} {\bibfnamefont {Y.~G.}\ \bibnamefont {Yao}},\
  }\bibfield  {title} {\enquote {\bibinfo {title} {Quantum spin {H}all effect
  in silicene and two-dimensional germanium},}\ }\href {\doibase
  10.1103/PhysRevLett.107.076802} {\bibfield  {journal} {\bibinfo  {journal}
  {Phys. Rev. Lett.}\ }\textbf {\bibinfo {volume} {107}},\ \bibinfo {pages}
  {076802} (\bibinfo {year} {2011})}\BibitemShut {NoStop}%
\bibitem [{\citenamefont {Lu}\ \emph {et~al.}(2016)\citenamefont {Lu},
  \citenamefont {Zhou}, \citenamefont {Chang}, \citenamefont {Guan},
  \citenamefont {Chen}, \citenamefont {Jiang}, \citenamefont {Jiang},
  \citenamefont {Wang}, \citenamefont {Yang}, \citenamefont {Feng},
  \citenamefont {dKawazoe},\ and\ \citenamefont {Lin}}]{Va-Dirac-LuYH}%
  \BibitemOpen
  \bibfield  {author} {\bibinfo {author} {\bibfnamefont {Y.~H.}\ \bibnamefont
  {Lu}}, \bibinfo {author} {\bibfnamefont {D.}~\bibnamefont {Zhou}}, \bibinfo
  {author} {\bibfnamefont {G.~Q.}\ \bibnamefont {Chang}}, \bibinfo {author}
  {\bibfnamefont {S.}~\bibnamefont {Guan}}, \bibinfo {author} {\bibfnamefont
  {W.~G.}\ \bibnamefont {Chen}}, \bibinfo {author} {\bibfnamefont {Y.~Z.}\
  \bibnamefont {Jiang}}, \bibinfo {author} {\bibfnamefont {J.~Z.}\ \bibnamefont
  {Jiang}}, \bibinfo {author} {\bibfnamefont {X.~S.}\ \bibnamefont {Wang}},
  \bibinfo {author} {\bibfnamefont {SY.~A}\ \bibnamefont {Yang}}, \bibinfo
  {author} {\bibfnamefont {Y.~P.}\ \bibnamefont {Feng}}, \bibinfo {author}
  {\bibfnamefont {Y.}~\bibnamefont {dKawazoe}}, \ and\ \bibinfo {author}
  {\bibfnamefont {H.}~\bibnamefont {Lin}},\ }\bibfield  {title} {\enquote
  {\bibinfo {title} {Multiple unpinned {D}irac points in group-{V}a
  single-layers with phosphorene structure},}\ }\href {\doibase
  10.1038/npjcompumats.2016.11} {\bibfield  {journal} {\bibinfo  {journal}
  {Npj. Comput. Mater.}\ }\textbf {\bibinfo {volume} {2}},\ \bibinfo {pages}
  {16011} (\bibinfo {year} {2016})}\BibitemShut {NoStop}%
\bibitem [{\citenamefont {Lee}\ \emph {et~al.}(2013)\citenamefont {Lee},
  \citenamefont {Kim}, \citenamefont {Toda}, \citenamefont {Matsuishi},\ and\
  \citenamefont {Hosono}}]{Nature-Ca2N-Hosono}%
  \BibitemOpen
  \bibfield  {author} {\bibinfo {author} {\bibfnamefont {K.}~\bibnamefont
  {Lee}}, \bibinfo {author} {\bibfnamefont {S.~W.}\ \bibnamefont {Kim}},
  \bibinfo {author} {\bibfnamefont {Y.}~\bibnamefont {Toda}}, \bibinfo {author}
  {\bibfnamefont {S.}~\bibnamefont {Matsuishi}}, \ and\ \bibinfo {author}
  {\bibfnamefont {H.}~\bibnamefont {Hosono}},\ }\bibfield  {title} {\enquote
  {\bibinfo {title} {Dicalcium nitride as a two-dimensional electride with an
  anionic electron layer},}\ }\href {\doibase 10.1038/nature11812} {\bibfield
  {journal} {\bibinfo  {journal} {Nature}\ }\textbf {\bibinfo {volume} {494}},\
  \bibinfo {pages} {336--340} (\bibinfo {year} {2013})}\BibitemShut {NoStop}%
\bibitem [{\citenamefont {Khazaei}\ \emph {et~al.}(2013)\citenamefont
  {Khazaei}, \citenamefont {Arai}, \citenamefont {Sasaki}, \citenamefont
  {Chung}, \citenamefont {Venkataramanan}, \citenamefont {Estili},
  \citenamefont {Sakka},\ and\ \citenamefont {Kawazoe}}]{AFM-MXene13-Kawazoe}%
  \BibitemOpen
  \bibfield  {author} {\bibinfo {author} {\bibfnamefont {M.}~\bibnamefont
  {Khazaei}}, \bibinfo {author} {\bibfnamefont {M.}~\bibnamefont {Arai}},
  \bibinfo {author} {\bibfnamefont {T.}~\bibnamefont {Sasaki}}, \bibinfo
  {author} {\bibfnamefont {C.}~\bibnamefont {Chung}}, \bibinfo {author}
  {\bibfnamefont {N.~S.}\ \bibnamefont {Venkataramanan}}, \bibinfo {author}
  {\bibfnamefont {M.}~\bibnamefont {Estili}}, \bibinfo {author} {\bibfnamefont
  {Y.}~\bibnamefont {Sakka}}, \ and\ \bibinfo {author} {\bibfnamefont
  {Y.}~\bibnamefont {Kawazoe}},\ }\bibfield  {title} {\enquote {\bibinfo
  {title} {Novel electronic and magnetic properties of two‐dimensional
  transition metal carbides and nitrides},}\ }\href {\doibase
  10.1002/adfm.201202502} {\bibfield  {journal} {\bibinfo  {journal} {Adv.
  Funct. Mater.}\ }\textbf {\bibinfo {volume} {23}},\ \bibinfo {pages}
  {2185--2192} (\bibinfo {year} {2013})}\BibitemShut {NoStop}%
\bibitem [{\citenamefont {Bhimanapati}\ \emph {et~al.}(2015)\citenamefont
  {Bhimanapati}, \citenamefont {Lin}, \citenamefont {Meunier}, \citenamefont
  {Jung}, \citenamefont {Cha}, \citenamefont {Das}, \citenamefont {Xiao},
  \citenamefont {Son}, \citenamefont {Strano}, \citenamefont {Cooper},
  \citenamefont {Liang}, \citenamefont {Louie}, \citenamefont {Ringe},
  \citenamefont {Kim}, \citenamefont {Naik}, \citenamefont {Sumpter},
  \citenamefont {Terrones}, \citenamefont {Xia}, \citenamefont {Wang},
  \citenamefont {Zhu}, \citenamefont {Akinwande}, \citenamefont {Alem},
  \citenamefont {Schuller}, \citenamefont {Schaak}, \citenamefont {Terrones},\
  and\ \citenamefont {Robinson}}]{2Dreview15_ACSnano_Robinson}%
  \BibitemOpen
  \bibfield  {author} {\bibinfo {author} {\bibfnamefont {G.~R.}\ \bibnamefont
  {Bhimanapati}}, \bibinfo {author} {\bibfnamefont {Z.}~\bibnamefont {Lin}},
  \bibinfo {author} {\bibfnamefont {V.}~\bibnamefont {Meunier}}, \bibinfo
  {author} {\bibfnamefont {Y.}~\bibnamefont {Jung}}, \bibinfo {author}
  {\bibfnamefont {J.}~\bibnamefont {Cha}}, \bibinfo {author} {\bibfnamefont
  {S.}~\bibnamefont {Das}}, \bibinfo {author} {\bibfnamefont {D.}~\bibnamefont
  {Xiao}}, \bibinfo {author} {\bibfnamefont {Y.}~\bibnamefont {Son}}, \bibinfo
  {author} {\bibfnamefont {M.~S.}\ \bibnamefont {Strano}}, \bibinfo {author}
  {\bibfnamefont {V.~R.}\ \bibnamefont {Cooper}}, \bibinfo {author}
  {\bibfnamefont {L.}~\bibnamefont {Liang}}, \bibinfo {author} {\bibfnamefont
  {S.~G.}\ \bibnamefont {Louie}}, \bibinfo {author} {\bibfnamefont
  {W.}~\bibnamefont {Ringe}, \bibfnamefont {E.~andZhou}}, \bibinfo {author}
  {\bibfnamefont {S.~S.}\ \bibnamefont {Kim}}, \bibinfo {author} {\bibfnamefont
  {R.~R.}\ \bibnamefont {Naik}}, \bibinfo {author} {\bibfnamefont {B.~G.}\
  \bibnamefont {Sumpter}}, \bibinfo {author} {\bibfnamefont {H.}~\bibnamefont
  {Terrones}}, \bibinfo {author} {\bibfnamefont {F.}~\bibnamefont {Xia}},
  \bibinfo {author} {\bibfnamefont {Y.}~\bibnamefont {Wang}}, \bibinfo {author}
  {\bibfnamefont {J.}~\bibnamefont {Zhu}}, \bibinfo {author} {\bibfnamefont
  {D.}~\bibnamefont {Akinwande}}, \bibinfo {author} {\bibfnamefont
  {N.}~\bibnamefont {Alem}}, \bibinfo {author} {\bibfnamefont {J.~A.}\
  \bibnamefont {Schuller}}, \bibinfo {author} {\bibfnamefont {R.~E.}\
  \bibnamefont {Schaak}}, \bibinfo {author} {\bibfnamefont {M.}~\bibnamefont
  {Terrones}}, \ and\ \bibinfo {author} {\bibfnamefont {J.~A.}\ \bibnamefont
  {Robinson}},\ }\bibfield  {title} {\enquote {\bibinfo {title} {Recent
  advances in two-dimensional materials beyond graphene},}\ }\href {\doibase
  10.1021/acsnano.5b05556} {\bibfield  {journal} {\bibinfo  {journal} {ACS
  Nano}\ }\textbf {\bibinfo {volume} {9}},\ \bibinfo {pages} {11509--11539}
  (\bibinfo {year} {2015})}\BibitemShut {NoStop}%
\bibitem [{\citenamefont {Geim}\ and\ \citenamefont
  {Grigorieva}(2013)}]{Geim13Nature_vdWCrystals}%
  \BibitemOpen
  \bibfield  {author} {\bibinfo {author} {\bibfnamefont {A.~K.}\ \bibnamefont
  {Geim}}\ and\ \bibinfo {author} {\bibfnamefont {I.~V.}\ \bibnamefont
  {Grigorieva}},\ }\bibfield  {title} {\enquote {\bibinfo {title} {Van der
  {Waals} heterostructures},}\ }\href {\doibase 10.1038/nature12385} {\bibfield
   {journal} {\bibinfo  {journal} {Nature}\ }\textbf {\bibinfo {volume}
  {499}},\ \bibinfo {pages} {419--425} (\bibinfo {year} {2013})}\BibitemShut
  {NoStop}%
\bibitem [{\citenamefont {Mak}\ \emph {et~al.}(2010)\citenamefont {Mak},
  \citenamefont {Lee}, \citenamefont {Hone}, \citenamefont {Shan},\ and\
  \citenamefont {Heinz}}]{MoS2-PRL-Heinz}%
  \BibitemOpen
  \bibfield  {author} {\bibinfo {author} {\bibfnamefont {K.~F.}\ \bibnamefont
  {Mak}}, \bibinfo {author} {\bibfnamefont {C.}~\bibnamefont {Lee}}, \bibinfo
  {author} {\bibfnamefont {J.}~\bibnamefont {Hone}}, \bibinfo {author}
  {\bibfnamefont {J.}~\bibnamefont {Shan}}, \ and\ \bibinfo {author}
  {\bibfnamefont {T.~F.}\ \bibnamefont {Heinz}},\ }\bibfield  {title} {\enquote
  {\bibinfo {title} {Atomically thin {M}o{S}$_{2}$: A new direct-gap
  semiconductor},}\ }\href {\doibase 10.1103/PhysRevLett.105.136805} {\bibfield
   {journal} {\bibinfo  {journal} {Phys. Rev. Lett.}\ }\textbf {\bibinfo
  {volume} {105}},\ \bibinfo {pages} {136805} (\bibinfo {year}
  {2010})}\BibitemShut {NoStop}%
\bibitem [{\citenamefont {Radisavljevic}\ \emph {et~al.}(2011)\citenamefont
  {Radisavljevic}, \citenamefont {Radenovic}, \citenamefont {Brivio},
  \citenamefont {Giacometti},\ and\ \citenamefont
  {Kis}}]{MoS2-FETmobility-NatNano-Kis}%
  \BibitemOpen
  \bibfield  {author} {\bibinfo {author} {\bibfnamefont {B.}~\bibnamefont
  {Radisavljevic}}, \bibinfo {author} {\bibfnamefont {A.}~\bibnamefont
  {Radenovic}}, \bibinfo {author} {\bibfnamefont {J.}~\bibnamefont {Brivio}},
  \bibinfo {author} {\bibfnamefont {V.}~\bibnamefont {Giacometti}}, \ and\
  \bibinfo {author} {\bibfnamefont {A.}~\bibnamefont {Kis}},\ }\bibfield
  {title} {\enquote {\bibinfo {title} {Single-layer {M}o{S}$_{2}$
  transistors},}\ }\href {\doibase 10.1038/nnano.2010.279} {\bibfield
  {journal} {\bibinfo  {journal} {Nature Nanotech.}\ }\textbf {\bibinfo
  {volume} {6}},\ \bibinfo {pages} {147--150} (\bibinfo {year}
  {2011})}\BibitemShut {NoStop}%
\bibitem [{\citenamefont {Zheng}\ \emph {et~al.}(2015)\citenamefont {Zheng},
  \citenamefont {Chen}, \citenamefont {Ng}, \citenamefont {Xu}, \citenamefont
  {Liu}, \citenamefont {Li}, \citenamefont {O'Shea}, \citenamefont
  {Dumitric\u{a}},\ and\ \citenamefont {Loh}}]{MoS2-Rippling-PRL15-ZY}%
  \BibitemOpen
  \bibfield  {author} {\bibinfo {author} {\bibfnamefont {Y.}~\bibnamefont
  {Zheng}}, \bibinfo {author} {\bibfnamefont {J.}~\bibnamefont {Chen}},
  \bibinfo {author} {\bibfnamefont {M.~F.}\ \bibnamefont {Ng}}, \bibinfo
  {author} {\bibfnamefont {H.}~\bibnamefont {Xu}}, \bibinfo {author}
  {\bibfnamefont {Y.~P.}\ \bibnamefont {Liu}}, \bibinfo {author} {\bibfnamefont
  {A.}~\bibnamefont {Li}}, \bibinfo {author} {\bibfnamefont {S.~J.}\
  \bibnamefont {O'Shea}}, \bibinfo {author} {\bibfnamefont {T.}~\bibnamefont
  {Dumitric\u{a}}}, \ and\ \bibinfo {author} {\bibfnamefont {K.~P.}\
  \bibnamefont {Loh}},\ }\bibfield  {title} {\enquote {\bibinfo {title}
  {Quantum mechanical rippling of a {M}o{S}$_{2}$ monolayer controlled by
  interlayer bilayer coupling},}\ }\href {\doibase
  10.1103/PhysRevLett.114.065501} {\bibfield  {journal} {\bibinfo  {journal}
  {Phys. Rev. Lett.}\ }\textbf {\bibinfo {volume} {114}},\ \bibinfo {pages}
  {065501} (\bibinfo {year} {2015})}\BibitemShut {NoStop}%
\bibitem [{\citenamefont {Wang}\ \emph
  {et~al.}(2018{\natexlab{a}})\citenamefont {Wang}, \citenamefont {Chernikov},
  \citenamefont {Glazov}, \citenamefont {Heinz}, \citenamefont {Marie},
  \citenamefont {Amand},\ and\ \citenamefont
  {Urbaszek}}]{Excitons-TMDCs-ROM18-Heinz}%
  \BibitemOpen
  \bibfield  {author} {\bibinfo {author} {\bibfnamefont {G.}~\bibnamefont
  {Wang}}, \bibinfo {author} {\bibfnamefont {A.}~\bibnamefont {Chernikov}},
  \bibinfo {author} {\bibfnamefont {M.~M.}\ \bibnamefont {Glazov}}, \bibinfo
  {author} {\bibfnamefont {T.~F.}\ \bibnamefont {Heinz}}, \bibinfo {author}
  {\bibfnamefont {X.}~\bibnamefont {Marie}}, \bibinfo {author} {\bibfnamefont
  {T.}~\bibnamefont {Amand}}, \ and\ \bibinfo {author} {\bibfnamefont
  {B.}~\bibnamefont {Urbaszek}},\ }\bibfield  {title} {\enquote {\bibinfo
  {title} {Colloquium: Excitons in atomically thin transition metal
  dichalcogenides},}\ }\href {\doibase 10.1103/RevModPhys.90.021001} {\bibfield
   {journal} {\bibinfo  {journal} {Rev. Mod. Phys.}\ }\textbf {\bibinfo
  {volume} {90}},\ \bibinfo {pages} {021001} (\bibinfo {year}
  {2018}{\natexlab{a}})}\BibitemShut {NoStop}%
\bibitem [{\citenamefont {Li}\ \emph {et~al.}(2014)\citenamefont {Li},
  \citenamefont {Yu}, \citenamefont {Ye}, \citenamefont {Ge}, \citenamefont
  {Ou}, \citenamefont {Wu}, \citenamefont {Feng}, \citenamefont {Chen},\ and\
  \citenamefont {Zhang}}]{BP-FET-ZhangYB-NatNano}%
  \BibitemOpen
  \bibfield  {author} {\bibinfo {author} {\bibfnamefont {L.}~\bibnamefont
  {Li}}, \bibinfo {author} {\bibfnamefont {Y.}~\bibnamefont {Yu}}, \bibinfo
  {author} {\bibfnamefont {G.~J.}\ \bibnamefont {Ye}}, \bibinfo {author}
  {\bibfnamefont {Q.}~\bibnamefont {Ge}}, \bibinfo {author} {\bibfnamefont
  {X.}~\bibnamefont {Ou}}, \bibinfo {author} {\bibfnamefont {H.}~\bibnamefont
  {Wu}}, \bibinfo {author} {\bibfnamefont {D.}~\bibnamefont {Feng}}, \bibinfo
  {author} {\bibfnamefont {X.~H.}\ \bibnamefont {Chen}}, \ and\ \bibinfo
  {author} {\bibfnamefont {Y.~B.}\ \bibnamefont {Zhang}},\ }\bibfield  {title}
  {\enquote {\bibinfo {title} {Black phosphorus field-effect transistors},}\
  }\href {\doibase 10.1038/nnano.2014.35} {\bibfield  {journal} {\bibinfo
  {journal} {Nature Nanotech.}\ }\textbf {\bibinfo {volume} {9}},\ \bibinfo
  {pages} {372--377} (\bibinfo {year} {2014})}\BibitemShut {NoStop}%
\bibitem [{\citenamefont {Li}\ \emph {et~al.}(2016)\citenamefont {Li},
  \citenamefont {Yang}, \citenamefont {Ye}, \citenamefont {Zhang},
  \citenamefont {Zhu}, \citenamefont {Lou}, \citenamefont {Zhou}, \citenamefont
  {Li}, \citenamefont {Watanabe}, \citenamefont {Taniguchi}, \citenamefont
  {Chang}, \citenamefont {Wang}, \citenamefont {Chen},\ and\ \citenamefont
  {Zhang}}]{ZhangYB_16BPQHE_NatNano}%
  \BibitemOpen
  \bibfield  {author} {\bibinfo {author} {\bibfnamefont {L.~K.}\ \bibnamefont
  {Li}}, \bibinfo {author} {\bibfnamefont {F.~Y.}\ \bibnamefont {Yang}},
  \bibinfo {author} {\bibfnamefont {G.~J.}\ \bibnamefont {Ye}}, \bibinfo
  {author} {\bibfnamefont {Z.~C.}\ \bibnamefont {Zhang}}, \bibinfo {author}
  {\bibfnamefont {Z.~W.}\ \bibnamefont {Zhu}}, \bibinfo {author} {\bibfnamefont
  {W.~K.}\ \bibnamefont {Lou}}, \bibinfo {author} {\bibfnamefont {X.~Y.}\
  \bibnamefont {Zhou}}, \bibinfo {author} {\bibfnamefont {L.}~\bibnamefont
  {Li}}, \bibinfo {author} {\bibfnamefont {K.}~\bibnamefont {Watanabe}},
  \bibinfo {author} {\bibfnamefont {T.}~\bibnamefont {Taniguchi}}, \bibinfo
  {author} {\bibfnamefont {K.}~\bibnamefont {Chang}}, \bibinfo {author}
  {\bibfnamefont {Y.~Y.}\ \bibnamefont {Wang}}, \bibinfo {author}
  {\bibfnamefont {X.~H.}\ \bibnamefont {Chen}}, \ and\ \bibinfo {author}
  {\bibfnamefont {Y.~B.}\ \bibnamefont {Zhang}},\ }\bibfield  {title} {\enquote
  {\bibinfo {title} {Quantum {H}all effect in black phosphorus two-dimensional
  electron system},}\ }\href {\doibase 10.1038/NNANO.2016.42} {\bibfield
  {journal} {\bibinfo  {journal} {Nature Nanotech.}\ }\textbf {\bibinfo
  {volume} {11}},\ \bibinfo {pages} {592--596} (\bibinfo {year}
  {2016})}\BibitemShut {NoStop}%
\bibitem [{\citenamefont {Cao}\ \emph {et~al.}(2015)\citenamefont {Cao},
  \citenamefont {Li},\ and\ \citenamefont {Louie}}]{GaSe_GateFM_PRL}%
  \BibitemOpen
  \bibfield  {author} {\bibinfo {author} {\bibfnamefont {T.}~\bibnamefont
  {Cao}}, \bibinfo {author} {\bibfnamefont {Z.}~\bibnamefont {Li}}, \ and\
  \bibinfo {author} {\bibfnamefont {S.~G.}\ \bibnamefont {Louie}},\ }\bibfield
  {title} {\enquote {\bibinfo {title} {Tunable magnetism and half-metallicity
  in hole-doped monolayer {G}a{S}e},}\ }\href {\doibase
  10.1103/PhysRevLett.114.236602} {\bibfield  {journal} {\bibinfo  {journal}
  {Phys. Rev. Lett.}\ }\textbf {\bibinfo {volume} {114}},\ \bibinfo {pages}
  {236602} (\bibinfo {year} {2015})}\BibitemShut {NoStop}%
\bibitem [{\citenamefont {Banhurin}\ \emph {et~al.}(2017)\citenamefont
  {Banhurin}, \citenamefont {Tyurnina}, \citenamefont {Yu}, \citenamefont
  {Mishchenko}, \citenamefont {Zolyomi}, \citenamefont {Morozov}, \citenamefont
  {Kumar}, \citenamefont {Gorbachev}, \citenamefont {Kudrynskyi}, \citenamefont
  {Pezzini}, \citenamefont {Kovalyuk}, \citenamefont {Zeitler}, \citenamefont
  {Novoselov}, \citenamefont {Patanè}, \citenamefont {Eaves}, \citenamefont
  {Grigorieva}, \citenamefont {Fal'ko}, \citenamefont {Geim},\ and\
  \citenamefont {Cao}}]{InSe_QHE_NatNano}%
  \BibitemOpen
  \bibfield  {author} {\bibinfo {author} {\bibfnamefont {D.~A.}\ \bibnamefont
  {Banhurin}}, \bibinfo {author} {\bibfnamefont {A.~V.}\ \bibnamefont
  {Tyurnina}}, \bibinfo {author} {\bibfnamefont {G.~L.}\ \bibnamefont {Yu}},
  \bibinfo {author} {\bibfnamefont {A.}~\bibnamefont {Mishchenko}}, \bibinfo
  {author} {\bibfnamefont {V.}~\bibnamefont {Zolyomi}}, \bibinfo {author}
  {\bibfnamefont {Sergey~V.}\ \bibnamefont {Morozov}}, \bibinfo {author}
  {\bibfnamefont {R.~K.}\ \bibnamefont {Kumar}}, \bibinfo {author}
  {\bibfnamefont {R.~V.}\ \bibnamefont {Gorbachev}}, \bibinfo {author}
  {\bibfnamefont {Z.~R.}\ \bibnamefont {Kudrynskyi}}, \bibinfo {author}
  {\bibfnamefont {S.}~\bibnamefont {Pezzini}}, \bibinfo {author} {\bibfnamefont
  {Z.~D.}\ \bibnamefont {Kovalyuk}}, \bibinfo {author} {\bibfnamefont
  {U.}~\bibnamefont {Zeitler}}, \bibinfo {author} {\bibfnamefont {K.~S.}\
  \bibnamefont {Novoselov}}, \bibinfo {author} {\bibfnamefont {A.}~\bibnamefont
  {Patanè}}, \bibinfo {author} {\bibfnamefont {L.}~\bibnamefont {Eaves}},
  \bibinfo {author} {\bibfnamefont {I.~V.}\ \bibnamefont {Grigorieva}},
  \bibinfo {author} {\bibfnamefont {V.~I.}\ \bibnamefont {Fal'ko}}, \bibinfo
  {author} {\bibfnamefont {A.~K.}\ \bibnamefont {Geim}}, \ and\ \bibinfo
  {author} {\bibfnamefont {Y.}~\bibnamefont {Cao}},\ }\bibfield  {title}
  {\enquote {\bibinfo {title} {High electron mobility, quantum {H}all effect
  and anomalous optical response in atomically thin {I}n{S}e},}\ }\href
  {\doibase 10.1038/nnano.2016.242} {\bibfield  {journal} {\bibinfo  {journal}
  {Nature Nanotech.}\ }\textbf {\bibinfo {volume} {12}},\ \bibinfo {pages}
  {223--227} (\bibinfo {year} {2017})}\BibitemShut {NoStop}%
\bibitem [{\citenamefont {Gong}\ \emph {et~al.}(2017)\citenamefont {Gong},
  \citenamefont {Li}, \citenamefont {Li}, \citenamefont {Ji}, \citenamefont
  {Stern}, \citenamefont {Xia}, \citenamefont {Cao}, \citenamefont {Bao},
  \citenamefont {Wang}, \citenamefont {Wang}, \citenamefont {Qiu},
  \citenamefont {Cava}, \citenamefont {Louie}, \citenamefont {Xia},\ and\
  \citenamefont {Zhang}}]{Cr2Ge2Te6_2DFM_Nature}%
  \BibitemOpen
  \bibfield  {author} {\bibinfo {author} {\bibfnamefont {C.}~\bibnamefont
  {Gong}}, \bibinfo {author} {\bibfnamefont {L.}~\bibnamefont {Li}}, \bibinfo
  {author} {\bibfnamefont {Z.}~\bibnamefont {Li}}, \bibinfo {author}
  {\bibfnamefont {H.}~\bibnamefont {Ji}}, \bibinfo {author} {\bibfnamefont
  {A.}~\bibnamefont {Stern}}, \bibinfo {author} {\bibfnamefont
  {Y.}~\bibnamefont {Xia}}, \bibinfo {author} {\bibfnamefont {T.}~\bibnamefont
  {Cao}}, \bibinfo {author} {\bibfnamefont {W.}~\bibnamefont {Bao}}, \bibinfo
  {author} {\bibfnamefont {C.}~\bibnamefont {Wang}}, \bibinfo {author}
  {\bibfnamefont {Y.}~\bibnamefont {Wang}}, \bibinfo {author} {\bibfnamefont
  {Z.~Q.}\ \bibnamefont {Qiu}}, \bibinfo {author} {\bibfnamefont {R.~J.}\
  \bibnamefont {Cava}}, \bibinfo {author} {\bibfnamefont {S.~G.}\ \bibnamefont
  {Louie}}, \bibinfo {author} {\bibfnamefont {J.}~\bibnamefont {Xia}}, \ and\
  \bibinfo {author} {\bibfnamefont {X.}~\bibnamefont {Zhang}},\ }\bibfield
  {title} {\enquote {\bibinfo {title} {Discovery of intrinsic ferromagnetism in
  two-dimensional van der {W}aals crystals},}\ }\href {\doibase
  10.1038/nature22060} {\bibfield  {journal} {\bibinfo  {journal} {Nature}\
  }\textbf {\bibinfo {volume} {546}},\ \bibinfo {pages} {265--269} (\bibinfo
  {year} {2017})}\BibitemShut {NoStop}%
\bibitem [{\citenamefont {Huang}\ \emph {et~al.}(2017)\citenamefont {Huang},
  \citenamefont {Clark}, \citenamefont {Navarro-Moratalla}, \citenamefont
  {Klein}, \citenamefont {Cheng}, \citenamefont {Seyler}, \citenamefont
  {Zhong}, \citenamefont {Schmidgall}, \citenamefont {McGuire}, \citenamefont
  {Cobden}, \citenamefont {Yao}, \citenamefont {Xiao}, \citenamefont
  {Jarillo-Herrero},\ and\ \citenamefont {Xu}}]{CrI3_2DFM_Nature}%
  \BibitemOpen
  \bibfield  {author} {\bibinfo {author} {\bibfnamefont {B.}~\bibnamefont
  {Huang}}, \bibinfo {author} {\bibfnamefont {G.}~\bibnamefont {Clark}},
  \bibinfo {author} {\bibfnamefont {E.}~\bibnamefont {Navarro-Moratalla}},
  \bibinfo {author} {\bibfnamefont {D.~R.}\ \bibnamefont {Klein}}, \bibinfo
  {author} {\bibfnamefont {R.}~\bibnamefont {Cheng}}, \bibinfo {author}
  {\bibfnamefont {K.~L.}\ \bibnamefont {Seyler}}, \bibinfo {author}
  {\bibfnamefont {D.}~\bibnamefont {Zhong}}, \bibinfo {author} {\bibfnamefont
  {E.}~\bibnamefont {Schmidgall}}, \bibinfo {author} {\bibfnamefont {M.~A.}\
  \bibnamefont {McGuire}}, \bibinfo {author} {\bibfnamefont {D.~H.}\
  \bibnamefont {Cobden}}, \bibinfo {author} {\bibfnamefont {W.}~\bibnamefont
  {Yao}}, \bibinfo {author} {\bibfnamefont {D.}~\bibnamefont {Xiao}}, \bibinfo
  {author} {\bibfnamefont {P.}~\bibnamefont {Jarillo-Herrero}}, \ and\ \bibinfo
  {author} {\bibfnamefont {X.}~\bibnamefont {Xu}},\ }\bibfield  {title}
  {\enquote {\bibinfo {title} {Layer-dependent ferromagnetism in a van der
  {W}aals crystal down to the monolayer limit},}\ }\href {\doibase
  10.1038/nature22391} {\bibfield  {journal} {\bibinfo  {journal} {Nature}\
  }\textbf {\bibinfo {volume} {546}},\ \bibinfo {pages} {270--273} (\bibinfo
  {year} {2017})}\BibitemShut {NoStop}%
\bibitem [{\citenamefont {Gomes}\ and\ \citenamefont
  {Carvalho}(2015)}]{SnSe_BinaryBP_PRB}%
  \BibitemOpen
  \bibfield  {author} {\bibinfo {author} {\bibfnamefont {L.~C.}\ \bibnamefont
  {Gomes}}\ and\ \bibinfo {author} {\bibfnamefont {A.}~\bibnamefont
  {Carvalho}},\ }\bibfield  {title} {\enquote {\bibinfo {title} {Phosphorene
  analogues: Isoelectronic two-dimensional group-{IV} monochalcogenides with
  orthorhombic structure},}\ }\href {\doibase 10.1103/PhysRevB.92.085406}
  {\bibfield  {journal} {\bibinfo  {journal} {Phys. Rev. B}\ }\textbf {\bibinfo
  {volume} {92}},\ \bibinfo {pages} {085406} (\bibinfo {year}
  {2015})}\BibitemShut {NoStop}%
\bibitem [{\citenamefont {Wu}\ and\ \citenamefont
  {Zeng}(2016)}]{SnSe_Multiferro_NL}%
  \BibitemOpen
  \bibfield  {author} {\bibinfo {author} {\bibfnamefont {M.}~\bibnamefont
  {Wu}}\ and\ \bibinfo {author} {\bibfnamefont {X.~C.}\ \bibnamefont {Zeng}},\
  }\bibfield  {title} {\enquote {\bibinfo {title} {Intrinsic ferroelasticity
  and/or multiferroicity in two-dimensional phosphorene and phosphorene
  analogues},}\ }\href {\doibase 10.1021/acs.nanolett.6b00726} {\bibfield
  {journal} {\bibinfo  {journal} {Nano Lett.}\ }\textbf {\bibinfo {volume}
  {16}},\ \bibinfo {pages} {3236--3241} (\bibinfo {year} {2016})}\BibitemShut
  {NoStop}%
\bibitem [{\citenamefont {Wang}\ \emph
  {et~al.}(2018{\natexlab{b}})\citenamefont {Wang}, \citenamefont {Fan},
  \citenamefont {Shen}, \citenamefont {Hua}, \citenamefont {Hu}, \citenamefont
  {Sheng}, \citenamefont {Lu}, \citenamefont {Fang}, \citenamefont {Qiu},
  \citenamefont {Lu}, \citenamefont {Liu}, \citenamefont {Liu}, \citenamefont
  {Huang}, \citenamefont {Xu}, \citenamefont {Shen},\ and\ \citenamefont
  {Zheng}}]{SnSe_SdH_NatCommun}%
  \BibitemOpen
  \bibfield  {author} {\bibinfo {author} {\bibfnamefont {Z.}~\bibnamefont
  {Wang}}, \bibinfo {author} {\bibfnamefont {C.}~\bibnamefont {Fan}}, \bibinfo
  {author} {\bibfnamefont {Z.~X.}\ \bibnamefont {Shen}}, \bibinfo {author}
  {\bibfnamefont {C.}~\bibnamefont {Hua}}, \bibinfo {author} {\bibfnamefont
  {Q.}~\bibnamefont {Hu}}, \bibinfo {author} {\bibfnamefont {F.}~\bibnamefont
  {Sheng}}, \bibinfo {author} {\bibfnamefont {Y.~H.}\ \bibnamefont {Lu}},
  \bibinfo {author} {\bibfnamefont {H.}~\bibnamefont {Fang}}, \bibinfo {author}
  {\bibfnamefont {Z.}~\bibnamefont {Qiu}}, \bibinfo {author} {\bibfnamefont
  {J.}~\bibnamefont {Lu}}, \bibinfo {author} {\bibfnamefont {Z.}~\bibnamefont
  {Liu}}, \bibinfo {author} {\bibfnamefont {W.}~\bibnamefont {Liu}}, \bibinfo
  {author} {\bibfnamefont {Y.}~\bibnamefont {Huang}}, \bibinfo {author}
  {\bibfnamefont {Z.~A.}\ \bibnamefont {Xu}}, \bibinfo {author} {\bibfnamefont
  {D.~W.}\ \bibnamefont {Shen}}, \ and\ \bibinfo {author} {\bibfnamefont
  {Y.}~\bibnamefont {Zheng}},\ }\bibfield  {title} {\enquote {\bibinfo {title}
  {Defects controlled hole doping and multivalley transport in {S}n{S}e single
  crystals},}\ }\href {\doibase 10.1038/s41467-017-02566-1} {\bibfield
  {journal} {\bibinfo  {journal} {Nature Commun.}\ }\textbf {\bibinfo {volume}
  {9}},\ \bibinfo {pages} {47} (\bibinfo {year}
  {2018}{\natexlab{b}})}\BibitemShut {NoStop}%
\bibitem [{\citenamefont {Naguib}\ \emph {et~al.}(2012)\citenamefont {Naguib},
  \citenamefont {Mashtalir}, \citenamefont {Carle}, \citenamefont {Presser},
  \citenamefont {Lu}, \citenamefont {Hultman}, \citenamefont {Gogotsi},\ and\
  \citenamefont {Barsoum}}]{ACSNano-Ti2C-Barsoum}%
  \BibitemOpen
  \bibfield  {author} {\bibinfo {author} {\bibfnamefont {M.}~\bibnamefont
  {Naguib}}, \bibinfo {author} {\bibfnamefont {O.}~\bibnamefont {Mashtalir}},
  \bibinfo {author} {\bibfnamefont {J.}~\bibnamefont {Carle}}, \bibinfo
  {author} {\bibfnamefont {V.}~\bibnamefont {Presser}}, \bibinfo {author}
  {\bibfnamefont {J.}~\bibnamefont {Lu}}, \bibinfo {author} {\bibfnamefont
  {L.}~\bibnamefont {Hultman}}, \bibinfo {author} {\bibfnamefont
  {Y.}~\bibnamefont {Gogotsi}}, \ and\ \bibinfo {author} {\bibfnamefont
  {M.~W.}\ \bibnamefont {Barsoum}},\ }\bibfield  {title} {\enquote {\bibinfo
  {title} {Two-dimensional transition metal carbides},}\ }\href {\doibase
  10.1021/nn204153h} {\bibfield  {journal} {\bibinfo  {journal} {ACS Nano}\
  }\textbf {\bibinfo {volume} {6}},\ \bibinfo {pages} {1322--1331} (\bibinfo
  {year} {2012})}\BibitemShut {NoStop}%
\bibitem [{\citenamefont {Tsai}\ \emph {et~al.}(1956)\citenamefont {Tsai},
  \citenamefont {Harris},\ and\ \citenamefont
  {Lassettre}}]{Cs2Oexp_Lassettre_JPC}%
  \BibitemOpen
  \bibfield  {author} {\bibinfo {author} {\bibfnamefont {K.}~\bibnamefont
  {Tsai}}, \bibinfo {author} {\bibfnamefont {P.~M.}\ \bibnamefont {Harris}}, \
  and\ \bibinfo {author} {\bibfnamefont {E.~N.}\ \bibnamefont {Lassettre}},\
  }\bibfield  {title} {\enquote {\bibinfo {title} {The crystal structure of
  cesium monoxide},}\ }\href {\doibase 10.1021/j150537a022} {\bibfield
  {journal} {\bibinfo  {journal} {J. Phys. Chem.}\ }\textbf {\bibinfo {volume}
  {60}},\ \bibinfo {pages} {338--344} (\bibinfo {year} {1956})}\BibitemShut
  {NoStop}%
\bibitem [{\citenamefont {Wilson}\ \emph {et~al.}(2001)\citenamefont {Wilson},
  \citenamefont {Di~Salvo},\ and\ \citenamefont
  {Mahajan}}]{TMDCsReview_AdvPhys}%
  \BibitemOpen
  \bibfield  {author} {\bibinfo {author} {\bibfnamefont {J.~A.}\ \bibnamefont
  {Wilson}}, \bibinfo {author} {\bibfnamefont {F.~J.}\ \bibnamefont
  {Di~Salvo}}, \ and\ \bibinfo {author} {\bibfnamefont {S.}~\bibnamefont
  {Mahajan}},\ }\bibfield  {title} {\enquote {\bibinfo {title} {Charge-density
  waves and superlattices in the metallic layered transition metal
  dichalcogenides},}\ }\href {\doibase 10.1080/00018737500101391} {\bibfield
  {journal} {\bibinfo  {journal} {Adv. Phys.}\ }\textbf {\bibinfo {volume}
  {50}},\ \bibinfo {pages} {1171--1248} (\bibinfo {year} {2001})}\BibitemShut
  {NoStop}%
\bibitem [{\citenamefont {Zacharia}\ \emph {et~al.}(2004)\citenamefont
  {Zacharia}, \citenamefont {Ulbricht},\ and\ \citenamefont
  {Hertel}}]{Gr-Cleavage-PRB}%
  \BibitemOpen
  \bibfield  {author} {\bibinfo {author} {\bibfnamefont {R.}~\bibnamefont
  {Zacharia}}, \bibinfo {author} {\bibfnamefont {H.}~\bibnamefont {Ulbricht}},
  \ and\ \bibinfo {author} {\bibfnamefont {T.}~\bibnamefont {Hertel}},\
  }\bibfield  {title} {\enquote {\bibinfo {title} {Interlayer cohesive energy
  of graphite from thermal desorption of polyaromatic hydrocarbons},}\ }\href
  {\doibase 10.1103/PhysRevB.69.155406} {\bibfield  {journal} {\bibinfo
  {journal} {Phys. Rev. B}\ }\textbf {\bibinfo {volume} {69}},\ \bibinfo
  {pages} {155406} (\bibinfo {year} {2004})}\BibitemShut {NoStop}%
\bibitem [{\citenamefont {Green}\ and\ \citenamefont
  {Keevers}(1995)}]{Si-absorb-exp}%
  \BibitemOpen
  \bibfield  {author} {\bibinfo {author} {\bibfnamefont {M.~A.}\ \bibnamefont
  {Green}}\ and\ \bibinfo {author} {\bibfnamefont {M.~J.}\ \bibnamefont
  {Keevers}},\ }\bibfield  {title} {\enquote {\bibinfo {title} {Optical
  properties of intrinsic silicon at 300 {K}},}\ }\href {\doibase
  10.1002/pip.4670030303} {\bibfield  {journal} {\bibinfo  {journal} {Prog.
  Photovolt: Res. Appl.}\ }\textbf {\bibinfo {volume} {3}},\ \bibinfo {pages}
  {189--192} (\bibinfo {year} {1995})}\BibitemShut {NoStop}%
\bibitem [{\citenamefont {Gofron}\ \emph {et~al.}(1994)\citenamefont {Gofron},
  \citenamefont {Campuzano}, \citenamefont {Abrikosov}, \citenamefont
  {Lindroos}, \citenamefont {Bansil}, \citenamefont {Ding}, \citenamefont
  {Koelling},\ and\ \citenamefont {Dabrowski}}]{YBCO-vanHove_PRL94}%
  \BibitemOpen
  \bibfield  {author} {\bibinfo {author} {\bibfnamefont {K.}~\bibnamefont
  {Gofron}}, \bibinfo {author} {\bibfnamefont {J.~C.}\ \bibnamefont
  {Campuzano}}, \bibinfo {author} {\bibfnamefont {A.~A.}\ \bibnamefont
  {Abrikosov}}, \bibinfo {author} {\bibfnamefont {M.}~\bibnamefont {Lindroos}},
  \bibinfo {author} {\bibfnamefont {A.}~\bibnamefont {Bansil}}, \bibinfo
  {author} {\bibfnamefont {H.}~\bibnamefont {Ding}}, \bibinfo {author}
  {\bibfnamefont {D.}~\bibnamefont {Koelling}}, \ and\ \bibinfo {author}
  {\bibfnamefont {B.}~\bibnamefont {Dabrowski}},\ }\bibfield  {title} {\enquote
  {\bibinfo {title} {Observation of an "extended" van {H}ove singularity in
  {YB}a$_{2}${C}u$_{4}${O}$_{8}$ by ultrahigh energy resolution angle-resolved
  photoemission},}\ }\href {\doibase 10.1103/PhysRevLett.73.3302} {\bibfield
  {journal} {\bibinfo  {journal} {Phys. Rev. Lett.}\ }\textbf {\bibinfo
  {volume} {73}},\ \bibinfo {pages} {3302} (\bibinfo {year}
  {1994})}\BibitemShut {NoStop}%
\bibitem [{\citenamefont {Qiao}\ \emph {et~al.}(2014)\citenamefont {Qiao},
  \citenamefont {Kong}, \citenamefont {Hu}, \citenamefont {Yang},\ and\
  \citenamefont {Ji}}]{BP-calculation-NC-JiW}%
  \BibitemOpen
  \bibfield  {author} {\bibinfo {author} {\bibfnamefont {J.}~\bibnamefont
  {Qiao}}, \bibinfo {author} {\bibfnamefont {X.}~\bibnamefont {Kong}}, \bibinfo
  {author} {\bibfnamefont {Z.}~\bibnamefont {Hu}}, \bibinfo {author}
  {\bibfnamefont {F.}~\bibnamefont {Yang}}, \ and\ \bibinfo {author}
  {\bibfnamefont {W.}~\bibnamefont {Ji}},\ }\bibfield  {title} {\enquote
  {\bibinfo {title} {High-mobility transport anisotropy and linear dichroism in
  few-layer black phosphorus},}\ }\href {\doibase 10.1038/ncomms5475}
  {\bibfield  {journal} {\bibinfo  {journal} {Nature Commun.}\ }\textbf
  {\bibinfo {volume} {5}},\ \bibinfo {pages} {4475} (\bibinfo {year}
  {2014})}\BibitemShut {NoStop}%
\bibitem [{\citenamefont {Ma}\ \emph {et~al.}(2017)\citenamefont {Ma},
  \citenamefont {Kuc},\ and\ \citenamefont {Heine}}]{Tl2O_JACS17_Heine}%
  \BibitemOpen
  \bibfield  {author} {\bibinfo {author} {\bibfnamefont {Y.~D.}\ \bibnamefont
  {Ma}}, \bibinfo {author} {\bibfnamefont {A.}~\bibnamefont {Kuc}}, \ and\
  \bibinfo {author} {\bibfnamefont {T.}~\bibnamefont {Heine}},\ }\bibfield
  {title} {\enquote {\bibinfo {title} {Single-layer {T}l$_{2}${O}: A
  metal-shrouded 2{D} semiconductor with high electronic mobility},}\ }\href
  {\doibase 10.1021/jacs.7b06296} {\bibfield  {journal} {\bibinfo  {journal}
  {J. Am. Chem. Soc.}\ }\textbf {\bibinfo {volume} {139}},\ \bibinfo {pages}
  {11694--11697} (\bibinfo {year} {2017})}\BibitemShut {NoStop}%
\bibitem [{\citenamefont {Peng}\ \emph {et~al.}(2009)\citenamefont {Peng},
  \citenamefont {Xiang}, \citenamefont {Wei}, \citenamefont {Li}, \citenamefont
  {Xia},\ and\ \citenamefont {Li}}]{ZnO-FM-PRL}%
  \BibitemOpen
  \bibfield  {author} {\bibinfo {author} {\bibfnamefont {H.}~\bibnamefont
  {Peng}}, \bibinfo {author} {\bibfnamefont {H.~J.}\ \bibnamefont {Xiang}},
  \bibinfo {author} {\bibfnamefont {S.~H.}\ \bibnamefont {Wei}}, \bibinfo
  {author} {\bibfnamefont {S.~S.}\ \bibnamefont {Li}}, \bibinfo {author}
  {\bibfnamefont {J.~B.}\ \bibnamefont {Xia}}, \ and\ \bibinfo {author}
  {\bibfnamefont {J.~B.}\ \bibnamefont {Li}},\ }\bibfield  {title} {\enquote
  {\bibinfo {title} {Origin and enhancement of hole-induced ferromagnetism in
  first-row ${d}^{0}$ semiconductors},}\ }\href {\doibase
  10.1103/PhysRevLett.102.017201} {\bibfield  {journal} {\bibinfo  {journal}
  {Phys. Rev. Lett.}\ }\textbf {\bibinfo {volume} {102}},\ \bibinfo {pages}
  {017201} (\bibinfo {year} {2009})}\BibitemShut {NoStop}%
\bibitem [{\citenamefont {Lei}\ \emph {et~al.}(2016)\citenamefont {Lei},
  \citenamefont {Cui}, \citenamefont {Xiang}, \citenamefont {Shang},
  \citenamefont {Wang}, \citenamefont {Ye}, \citenamefont {Luo}, \citenamefont
  {Wu}, \citenamefont {Sun},\ and\ \citenamefont
  {Chen}}]{FeSegate2016-PRL-ChenXH}%
  \BibitemOpen
  \bibfield  {author} {\bibinfo {author} {\bibfnamefont {B.}~\bibnamefont
  {Lei}}, \bibinfo {author} {\bibfnamefont {J.~H.}\ \bibnamefont {Cui}},
  \bibinfo {author} {\bibfnamefont {Z.~J.}\ \bibnamefont {Xiang}}, \bibinfo
  {author} {\bibfnamefont {C.}~\bibnamefont {Shang}}, \bibinfo {author}
  {\bibfnamefont {N.~Z.}\ \bibnamefont {Wang}}, \bibinfo {author}
  {\bibfnamefont {G.~J.}\ \bibnamefont {Ye}}, \bibinfo {author} {\bibfnamefont
  {X.~G.}\ \bibnamefont {Luo}}, \bibinfo {author} {\bibfnamefont
  {T.}~\bibnamefont {Wu}}, \bibinfo {author} {\bibfnamefont {Z.}~\bibnamefont
  {Sun}}, \ and\ \bibinfo {author} {\bibfnamefont {X.~H.}\ \bibnamefont
  {Chen}},\ }\bibfield  {title} {\enquote {\bibinfo {title} {Evolution of
  high-temperature superconductivity from a low-${T}_{c}$ phase tuned by
  carrier concentration in {F}e{S}e thin flakes},}\ }\href {\doibase
  10.1103/PhysRevLett.116.077002} {\bibfield  {journal} {\bibinfo  {journal}
  {Phys. Rev. Lett.}\ }\textbf {\bibinfo {volume} {116}},\ \bibinfo {pages}
  {077002} (\bibinfo {year} {2016})}\BibitemShut {NoStop}%
\end{thebibliography}
\end{document}